\PassOptionsToPackage{rgb}{xcolor}
\documentclass[conference]{IEEEtran}
\def\BibTeX{{\rm B\kern-.05em{\sc i\kern-.025em b}\kern-.08em
    T\kern-.1667em\lower.7ex\hbox{E}\kern-.125emX}}

\usepackage{IEEE_Preamble_JFreytes}

\makeatletter
\newcommand{\mathleft}{\@fleqntrue\@mathmargin0pt}
\newcommand{\mathcenter}{\@fleqnfalse}
\makeatother

\makeatletter
\renewcommand{\fnum@figure}{Fig. \thefigure}
\makeatother

\makeatletter
\IEEEoverridecommandlockouts
\graphicspath{ {Images/}}

\begin{document}

\title{Grid-Forming Control Based On Emulated Synchronous Condenser Strategy Compliant With Challenging Grid Code Requirements}
%
\author{\IEEEauthorblockN{
Julian Freytes,
Antoine Ross\'{e},
Valentin Costan and
Gr\'{e}goire Prime
\IEEEauthorblockA{{EDF R\&D, D\'{e}partement SYSTEME, 91120 Palaiseau, France}\\
{{Corresponding author: julian.freytes@edf.fr}}}
}\vspace{-0cm}}
\maketitle
%
%
\begin{abstract}
Future power systems will include high shares of inverter-based generation. There is a general consensus that for allowing this transition, the Grid-Forming (GFM) control approach would be of great value. This article presents a GFM control strategy which is based on the concept of an Emulated Synchronous Condenser in parallel with a controlled current source with an explicit representation of the swing equation. The advantage of this control is that it can cope with challenging grid code requirements such as severe phase jumps, balanced and unbalanced Fault Ride-Through (FRT), main grid disconnection and black start. All these scenarios can be surpassed with a single control structure with no further logic involved (e.g. fault detection to turn on or off different control parts, freezes, etc.). The proposed strategy is evaluated via time-domain simulations of a 2-MW Battery Energy Storage System (BESS). 
\end{abstract}
%
\begin{IEEEkeywords}
BESS, FRT, grid-forming control, current limitation, grid code requirements. 
\end{IEEEkeywords}
%
%
\section{Introduction}\label{sec:Intro}
%
\IEEEPARstart{T}{here} is a general consensus that Grid-Forming (GFM) control paradigm deployed in Voltage Source Converters (VSC) will play an important role in the massive inclusion of renewable energies in modern power systems \cite{ESIGgeneral,RossoGridForming, GsP_Cigre}. This control strategy, although not completely new, presents enormous advantages for the electrical grid: it can improve the overall stability and can provide valuable services such as black start, natural improvements and damping of grid frequency dynamics, and can be adapted to grids with variable Short-Circuit Levels (SCL), among others. 
\par In terms of applications, GFM inverters can be deployed in Battery Energy Storage Systems (BESS) \cite{ABB_AEMO}, HVDC \cite{KIT_VSM_FRT}, FACTS \cite{MITSUBISHI_FACTS_GFM}. The main differences between them in terms of behavior will be dictated by the energy source. 
\par From a systems standpoint, GFM inverters will also be required to cope with complex scenarios that are arising due to evolutions of the power grid. The strict requirements being introduced for these new equipment are becoming challenging for the converters, since it is demanded to withstand difficult grid scenarios like grid faults, phase jumps and extreme Rate of Change of Frequency (RoCoF) \cite{ESIGgeneral}. 
\par Even though the GFM converters provide comparable grid services to conventional generators, their main difference resides in overcurrent capabilities. Indeed, because of the thermal limits of semiconductors used in VSCs and cost-limiting sizing, GFM inverters generally have limited current headroom (between 5\% to 20\%), so it is imperative that they are equipped with mechanisms to limit the internal current in every situation to avoid damaging the equipment. However, since the vast majority of GFM converters resides on a Power-Frequency relation for synchronization, risk of instability is indeed a likely possibility while the current is being limited, as explained in \cite{GFInstability, Aalborg_VSM_Damp}.  For this reason, it is also of high importance to verify the transient stability of the system when the current is being limited {via control actions} and verify that the synchronization is maintained.
\par
\begin{figure}[!htbp] 
	\centering
	\includegraphics[width=0.99\columnwidth]{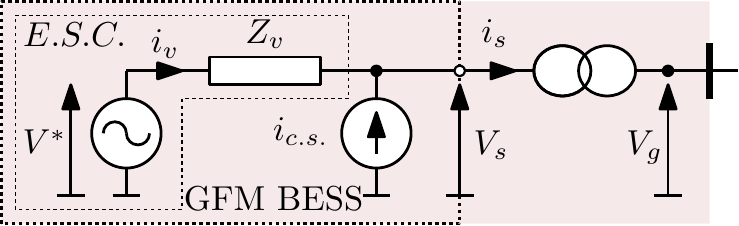}
	\caption{Emulated Synchronous Condenser (ESC) \cite{ABB_GFVC}.}
	\label{fig:EmSyncCond}
\end{figure}

 The contribution of this paper is the proposal of a Grid-Forming control strategy based on the concept of Emulated Synchronous Condenser (ESC) in parallel with a current source (illustrated in Fig.~\ref{fig:EmSyncCond}), which can be compliant with challenging grid code requirements. The approach of consi\-dering a virtual synchronous condenser was introduced in \cite{ABB_GFVC}, but relying on a Phase-Locked Loop (PLL) algorithm for the synchronization mechanism. Even though the authors highlight the parallel between the PLL gains and the swing equation, some assumptions were needed to arrive at the demonstration of this equivalence. Moreover, the difficulty of the PLL to ride through severe low-voltage faults requires the implementation of control modifications such as additional logic and freezing methods. {The approach proposed in this paper minimizes the necessity of logical decisions in the controller, hence, simplifying its implementation.}
\par In this paper, the synchronization is achieved by a classical swing equation, but instead of using the measured active power of the inverter, a virtual power is used as proposed in \cite{GFCurrLimSolution} for avoiding synchronization instabilities during current limitation. The difference in this article with respect to \cite{GFCurrLimSolution} resides in the fact that here, the negative sequence of voltages and currents are also considered, opening the possibilities for new degrees of freedom to be explored. {With this modification, fault ride-through of unbalanced scenarios can be accurately achieved.} 
%
%
\section{{Grid-Forming with Classical Virtual Synchronous Machine\label{sec:Classic_VSM}}} 
%
\par
\begin{figure}[!htbp] 
	\centering
	\includegraphics[width=0.99\columnwidth]{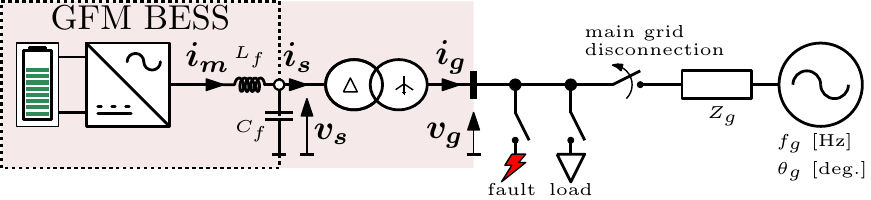}
	\caption{BESS Grid-forming system.}
	\label{fig:sys}
\end{figure}

 The considered system is shown in Fig.~\ref{fig:sys}, which is composed of a power-electronics (PE) stage, a sinus-filter, ac-coupling transformer and a specific grid configuration for each scenario. A two-level inverter is considered with a LCL output filter with explicit elements $L_f$ and $C_f$. The coupling transformer impedance provides the output inductance of the LCL \cite{AELeon_WF_FullConv}. The point of synchronization is given at the location of $C_f$, and the voltages and currents are given by $\bm{v_s}=[v_{sa},v_{sb},v_{sc}]^\top$ and $\bm{i_s}=[i_{sa},i_{sb},i_{sc}]^\top$, respectively.
\par
\begin{figure}[!htbp] 
	\centering
	\includegraphics[width=0.99\columnwidth]{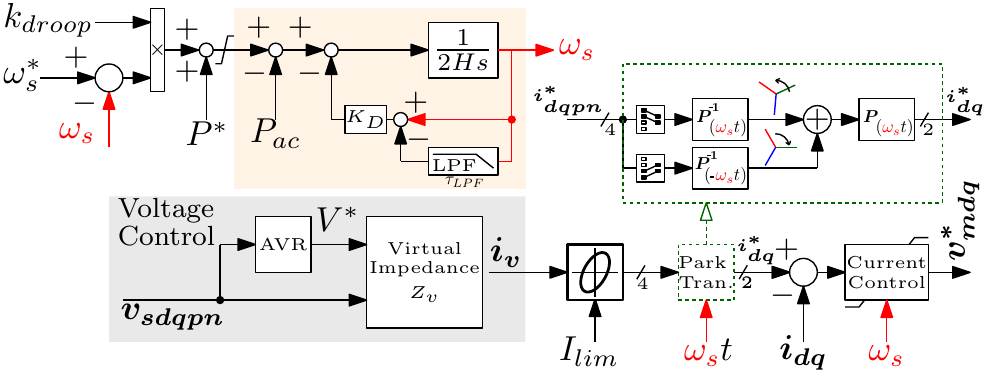}
	\caption{{Grid-Forming Control with Classical Virtual Synchronous Machine.}}
	\label{fig:GF_Ctrl_VSM}
\end{figure}

 {First, a classical GFM control based on the Virtual Synchronous Machine (VSM) is adopted as shown in Fig.~\ref{fig:GF_Ctrl_VSM}, which is detailed in the following.}
%
\subsection{{Virtual Synchronous Machine (VSM)}\label{sec:VSM}} 
\par The VSM implements the classical swing equation in per-unit given in \eqref{eq:VSM}, where $\ws=2\pi f_s$ is the rotating speed of the VSM; $H$ is the synthetic inertia constant; $K_D$ the damping factor and ``$1-LPF(s)$" a washout filter with $\tau_{LPF}=1$~ms. 
\begin{align}\label{eq:VSM}
	\frac{d\ws}{dt} = \frac{1}{2H} \left( P_{VSM}^* - P_{ac} - K_D \ws \left(1 - LPF(s)\right) \right)
\end{align}
\par {In this approach, the active power is controlled by accelerating the VSM as shown in \eqref{eq:VSM}, hence, the power dynamics will be linked to the parameters $H$, $K_D$ and $\tau_{LPF}$. The power $ P_{VSM}^*$ is obtained as the sum of the setpoint reference $P^*$ and the droop control $P_{droop}$, where:}
\begin{align}\label{eq:Pdroop}
	P_{droop} = k_{droop} \left( \omega_s^* - \ws \right) 
\end{align}
%
%
\subsection{Voltage Control and Virtual Impedance} \label{sec:VirtLine}
\par
\begin{figure}[!htbp] 
	\centering
	\includegraphics[width=0.99\columnwidth]{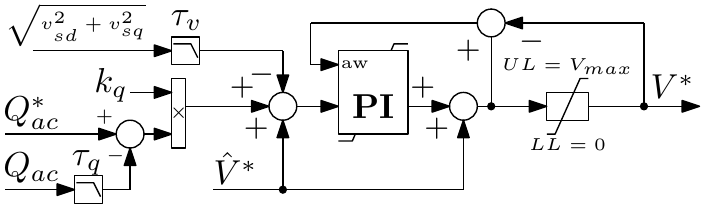}
	\caption{Voltage Controller with reactive power droop ``AVR'' in Fig.~\ref{fig:GF_Ctrl}.}
	\label{fig:avr}
\end{figure}

 The virtual impedance implemented in the controller is a quasi-stationary electrical model of a virtual impedance as explained in \cite{SINTEF_VSM_DQSM}. The current $\bm{i_v}$ is obtained as:
\begin{subequations}\label{Eq:QSEMdqpn}
	\begin{gather}
		\resizebox{0.85\hsize}{!}{$%
			\left[ 
			\begin{array}{c}
				i_{vdp}\\
				i_{vqp}\\
			\end{array}  
			\right]=
			\frac{1}{R_{vp}^2+X_{vp}^2}
			\left[
			\begin{array}{cc}
				R_{vp} & X_{vp}\\
				-X_{vp} & R_{vp}
			\end{array}
			\right]
			\left( 
			\begin{array}{c}
				V^*-v_{sdp}\\
				-v_{sqp}\\
			\end{array}
			\right)
			$}%
	\end{gather}
	\begin{gather}\label{Eq:QSEMdqn}
		\resizebox{0.85\hsize}{!}{$%
			\left[ 
			\begin{array}{c}
				i_{vdn}\\
				i_{vqn}\\
			\end{array}  
			\right]=
			\frac{1}{R_{vn}^2+X_{vn}^2}
			\left[
			\begin{array}{cc}
				R_{vn} & -X_{vn}\\
				X_{vn} & R_{vn}
			\end{array}
			\right]
			\left( 
			\begin{array}{c}
				-v_{sdn}\\
				-v_{sqn}\\
			\end{array}
			\right)
			$}%
	\end{gather}
\end{subequations}
\noindent where $V^*$ is the output of the Automatic Voltage Regulator (AVR) with reactive power droop as shown in Fig.~\ref{fig:avr} \cite{SINTEF_VSM_DQSM}. 

%
\subsection{Reference Current Limitation and Current Control\label{sec:CurrCtrl}} 
\par The {reference for the current controller is $\bm{i_v}$, which is the output of the virtual impedance block from \eqref{Eq:QSEMdqpn}}. Then it is limited by an elliptical limitation which consi\-ders the positive and negative sequences as explained in \cite{KevinGE}, obtaining the reference $\bm{i_{dqpn}^*}$. This method of direct current saturation provides a efficient mechanism for current limitation \cite{Taoufik2020}.
\par The current controller is implemented in $dq$ frame (2-axes) for avoiding the dynamics of the DDSRF in the feed-back path. For adapting the reference $\bm{i_{dqpn}}$ for the current controller, a combined Park transformation $\bm{P}$ is applied to couple both sequences in a positive-sequence $dq$ frame as in \cite{AELeon_WF_FullConv} {(note that the zero sequence current is zero due to the transformer connection)}. For obtaining $\bm{i_{dq}}$, Park transformation is applied to $\bm{i_m}=[i_{ma},i_{mb},i_{mc}]^\top$ (see Fig.~\ref{fig:sys}). Finally, the current controller is implemented with a classical PI controller {and resonant controllers acting at $2\ws$ for imbalances}. 
%
\subsection{{Instability of VSM during current limiting}\label{sec:VSM_FRT}} 
\par
\begin{figure}[t]
	\centering
	\begin{minipage}{1\columnwidth}
		\hfill
		{\includegraphics[trim={0 1.1cm 0 0},clip,width=0.987\columnwidth]{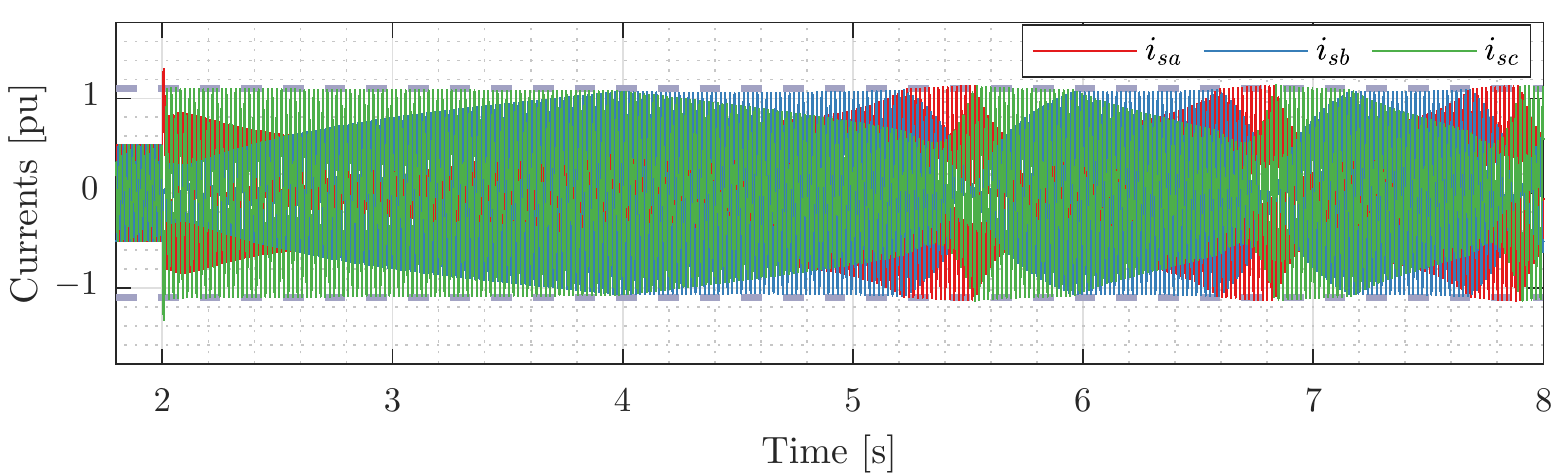}}
	\end{minipage}%
	\newline
	\begin{minipage}{1\columnwidth}
		\centering
		{\includegraphics[width=1\columnwidth]{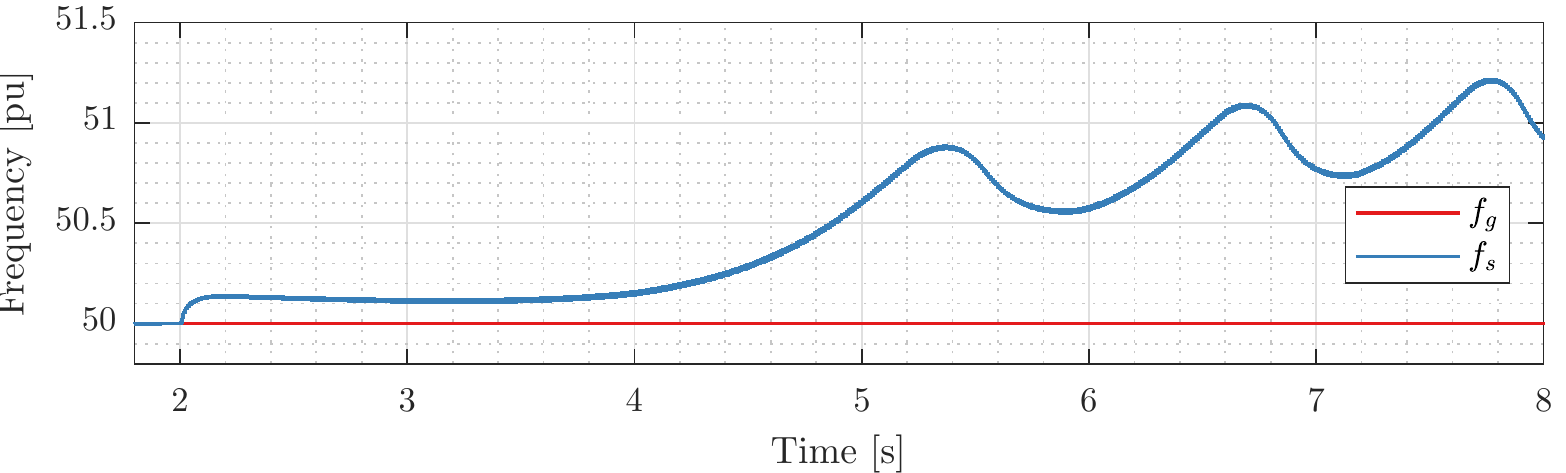}}
	\end{minipage}
	\newline
	\vspace{-0.2cm}
	\caption{{VSM Instability during current limitation --- Top:  ac currents in [p.u.] with $I_{base}^s$; Bottom: ac frequency in [Hz].}} \label{fig:VSM_FRT}
\end{figure}

 {For highlighting a major drawback of the VSM with current limitation}, the system shown in Fig.~\ref{fig:sys} is modelled in Matlab-Simulink considering a 2-MW BESS {and adopting the controller from Fig.~\ref{fig:GF_Ctrl_VSM}}. Main parameters are shown in Table~\ref{Table:BaseValues}. {Considering an initial power transmission of $P_{ac} = 0.5$~p.u., a single-phase fault is applied at $t=2$~s. Results of the ac currents and frequency of the VSM are shown in Fig.~\ref{fig:VSM_FRT}. Even if the fault is asymmetrical, the currents are controlled to avoid surpassing the limit imposed by $I_{lim}$ of $1.1$~p.u. However, the ac frequency is clearly unstable since the VSM cannot converge to a stable operation point. The virtual machine intends to provide higher currents but these are blocked by the current limitation method. Hence, the VSM augments the internal angle until the synchronisation is lost \cite{GFInstability}. Even if the instability is manifested after $3$~s and the faults would have been cleared by that time, the same instability may occur in case that the GFM inverter operates near its maximum power capabilities \cite{GsP_Cigre}.} 
\begin{table}[t!]
	\caption{System Parameters.}
	\vspace{-0.2cm}
	\centering
	\begin{tabular}{cccccc}
		\hline\hline 
		$S_{base}$  &  $=2000$~kVA        & $V_{base}^{g}$      & $=\sqrt{\frac{2}{3}} \;20$~kV  & $I_{base}^{g}$  & $=\frac{2}{3} \; \frac{S_{base}}{V_{base}^{g}}$~A \Tstrut\\ [1.5pt]
		$\omega_{base}$  &  $=2\pi 50$~rad/s               & $V_{base}^{s}$     & $=\sqrt{\frac{2}{3}} \;400$~V  & {$I_{base}^{s}$}   & $=\frac{2}{3} \; \frac{S_{base}}{V_{base}^{s}}$A \Tstrut\\ [1.5pt]
		$X_{tr}$ & $=6.5$~\% & $L_{f}$ & $=12$~\% & $C_f$ & $=5$~\% \Tstrut\\ [1.5pt]
		$H$ & $=2$~s & {$K_{D}$} & $=100\:s^{\minus 1}$ & $k_{droop}$ & $=1/0.05$~p.u. \Tstrut\\ [1.5pt]
		$L_{v}$ & $=0.2$~p.u. & $R_{v}$ & $=0.05$~p.u. & $I_{lim}$  & $=1.1$~p.u.  \Tstrut\\ [1.5pt]    
		$k_q$ & $=0.1$~p.u. & $\tau_{v}=\tau_{q}$ & $=5$~ms & $\tau_{LPF}$  & $=1$~ms  \Tstrut\\ [1.5pt]  \hline\hline
	\end{tabular}
	\label{Table:BaseValues}%
	\vspace{-0.5cm}
\end{table}%
%
\section{Grid-Forming Control with Emulated Synchronous Condenser\label{sec:Control}} 
%
%
\par
\begin{figure}[!htbp] 
	\centering
	\includegraphics[width=0.99\columnwidth]{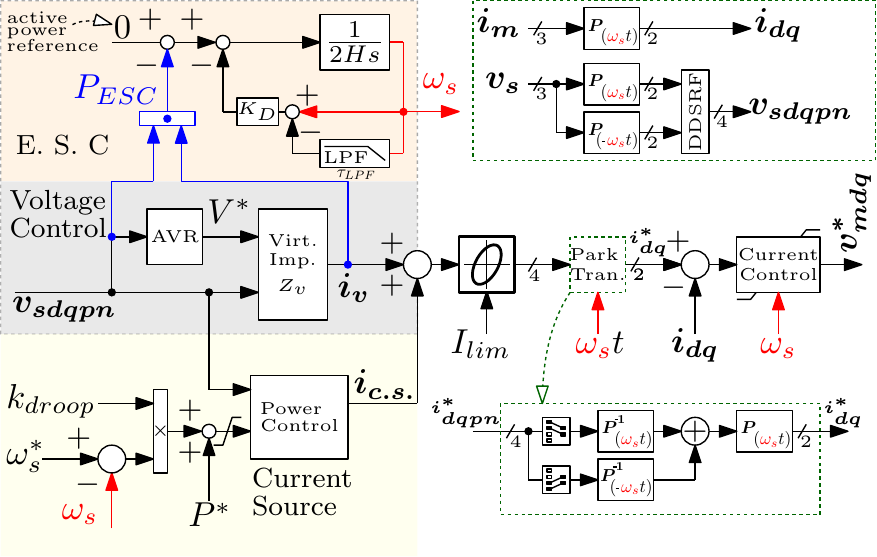}
    \caption{Grid-Forming Control with Emulated Synchronous Condenser.}   
	\label{fig:GF_Ctrl}
\end{figure}

 {To overcome the stability issue of the VSM, the control from Fig.~\ref{fig:GF_Ctrl_VSM} is modified as shown in} Fig.~\ref{fig:GF_Ctrl}. {The fundamental difference is that the synchronisation action part is obtained by an Emulated Synchronous Condenser (ESC), which is not intended to provide active power in normal operation. Instead, the active power is controlled by a current source which provides instantaneous current references to be tracked.} 
\subsection{Emulated Synchronous Condenser (ESC)\label{sec:Synch}} 
\par  {The ESC dynamics are similar to \eqref{eq:VSM}, but the power reference $P_{ESC}^*$ is always zero and the power $P_{ESC}$ is not the measured active power from the inverter, but an internal calculation given by ``$\bm{v_{sdqpn}} \bigcdot \bm{i_{v}}$''}. 
\begin{align}\label{eq:esc}
\resizebox{0.9\hsize}{!}{$%
\frac{d\ws}{dt} = \frac{1}{2H} \left( \underbrace{P_{ESC}^*}_{=0} - \underbrace{P_{ESC}}_{\bm{v_{sdqpn}} \bigcdot  \bm{i_{v}}} - K_D \ws \left(1 - LPF(s)\right) \right)
$}%
\end{align}
\subsection{Current Source: Active Power Control\label{sec:PQctrl}} 
\par For imposing the active power, the virtual current source illustrated in Fig.~\ref{fig:EmSyncCond} is used. The current $\bm{i_{c.s.}}=[i_{c.s.,dp},i_{c.s.,qp},0,0]^\top$ is obtained as:
\begin{subequations}\label{eq:currsource}
\begin{gather}
i_{c.s.,dp}= \left(P^* + P_{droop} \right)\frac{v_{sdp}}{v_{sdp}^2+v_{sqp}^2}  
\\
i_{c.s.,qp}=\left(P^* + P_{droop} \right)\frac{v_{sqp}}{v_{sdp}^2+v_{sqp}^2}
\end{gather}
\end{subequations}
\noindent Note that the negative sequence is set to zero, but this can be extended to provide active power through both sequences as in \cite{teodorescu2011grid} for example. The contribution from the Virtual Impedance $\bm{i_v}$ and the current source $\bm{i_{c.s.}}$ are added to form the total unsaturated current reference {as explained in Section~\ref{sec:CurrCtrl}}.
%

%
\section{Evaluation of GFM control for challenging requirements via time-domain simulations} \label{sec:Simulations}
%
\par {This section evaluates the behavior of the proposed control in different scenarios. The ac grid strength (i.e. SCL) is modified for the different scenarios to highlight the capabilities of the controller to cope with different SCLs.}
%
%
\subsection{Power step and power reversal\label{sec:Pstep}} 
\par
\begin{figure}[!htbp] 
	\centering
	\includegraphics[width=0.99\columnwidth]{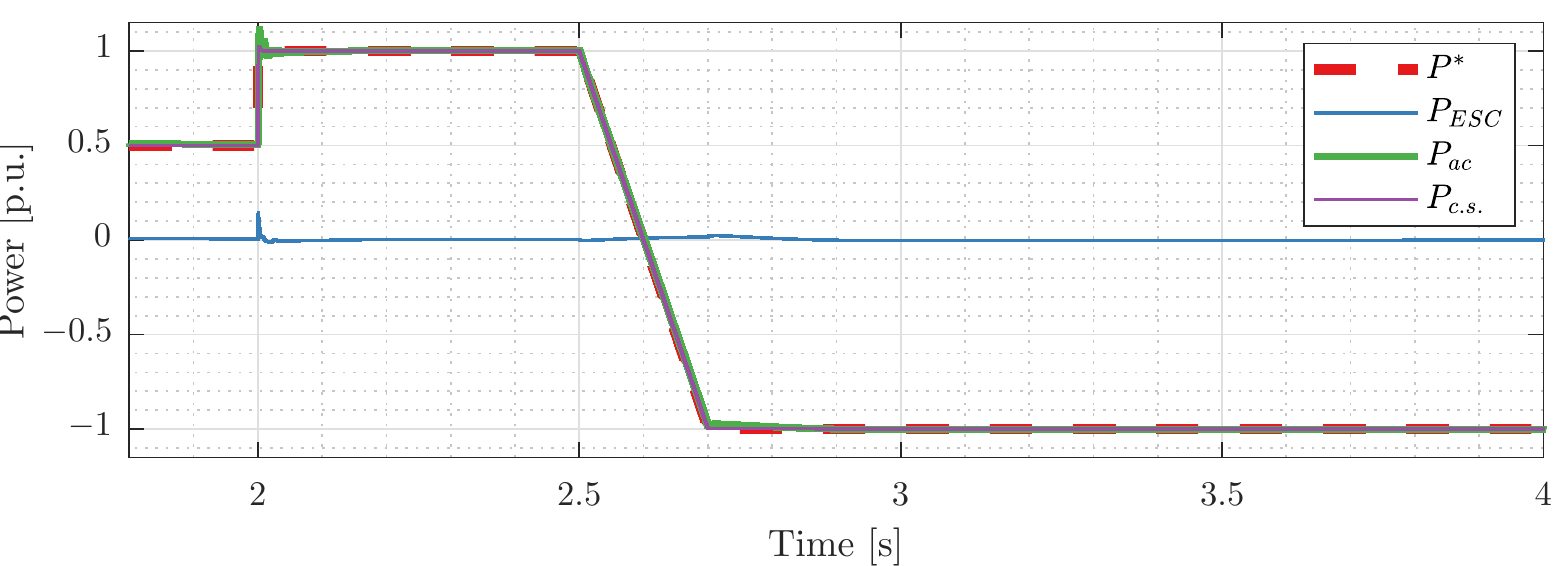}
	\vspace{-0.65cm}
	\caption{Active power reference step simulation results in [p.u.] with $S_{base}$.}
	\label{fig:Pstep_Power}
\end{figure}

 The first scenario consists in an active power step and fast power reversal in a strong ac grid (SCL = $400$~MVA). Starting from a power transfer of $0.5$~p.u., a step is applied at $t=2$~s up to $1$~p.u. Then, an active power reversal is applied at $t=2.5$~s. Results of the different active power responses are shown in Fig.~\ref{fig:Pstep_Power}. The power from current source $P_{c.s.}$ follows the reference $P^*$ accurately, but this power flows to the ac grid and also virtually to the emulated synchronous condenser, as shown by $P_{ESC}$. The sum of those signals is the real power flowing out of the inverter, $P_{ac}$. The decoupling can be improved by modifying dynamically the ac voltage reference from the virtual impedance function.

%
\subsection{Black start \label{sec:Blackstart}} 
\par
\begin{figure}[t]
	\centering
	\begin{minipage}{1\columnwidth}
		\centering
		{\includegraphics[trim={0 1.1cm 0 0},clip,width=1\columnwidth]{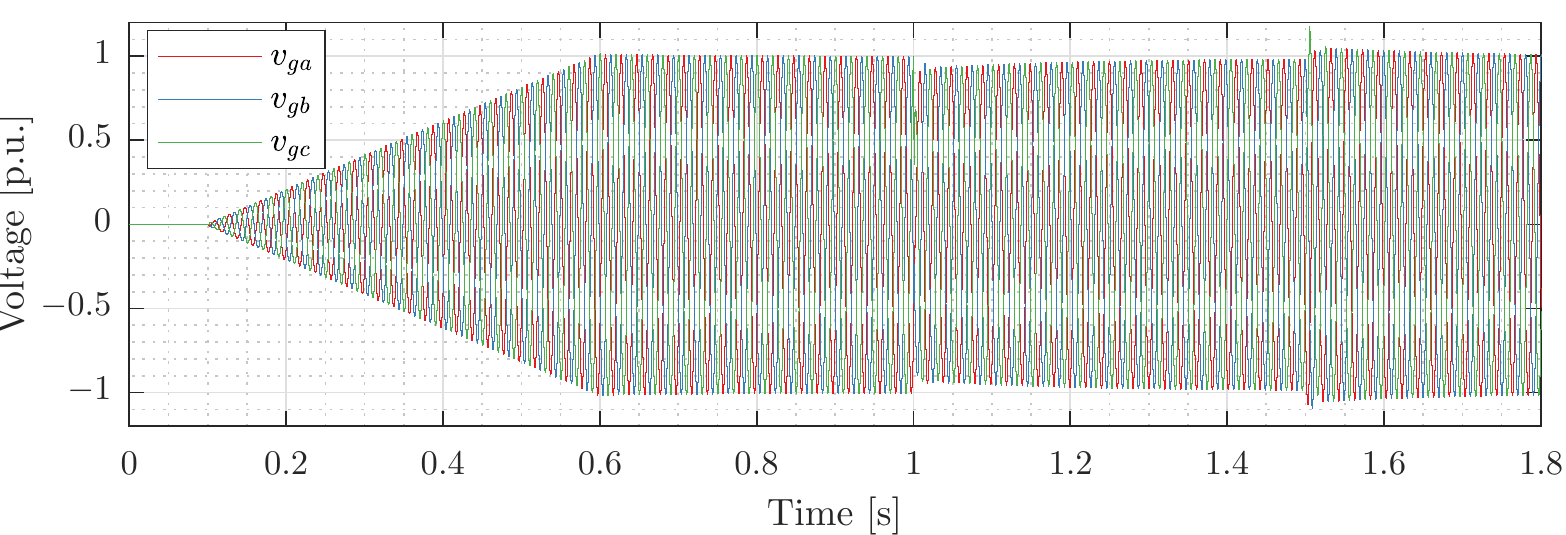}}
	\end{minipage}%
	\newline
	\begin{minipage}{1\columnwidth}
		\centering
		{\includegraphics[width=1\columnwidth]{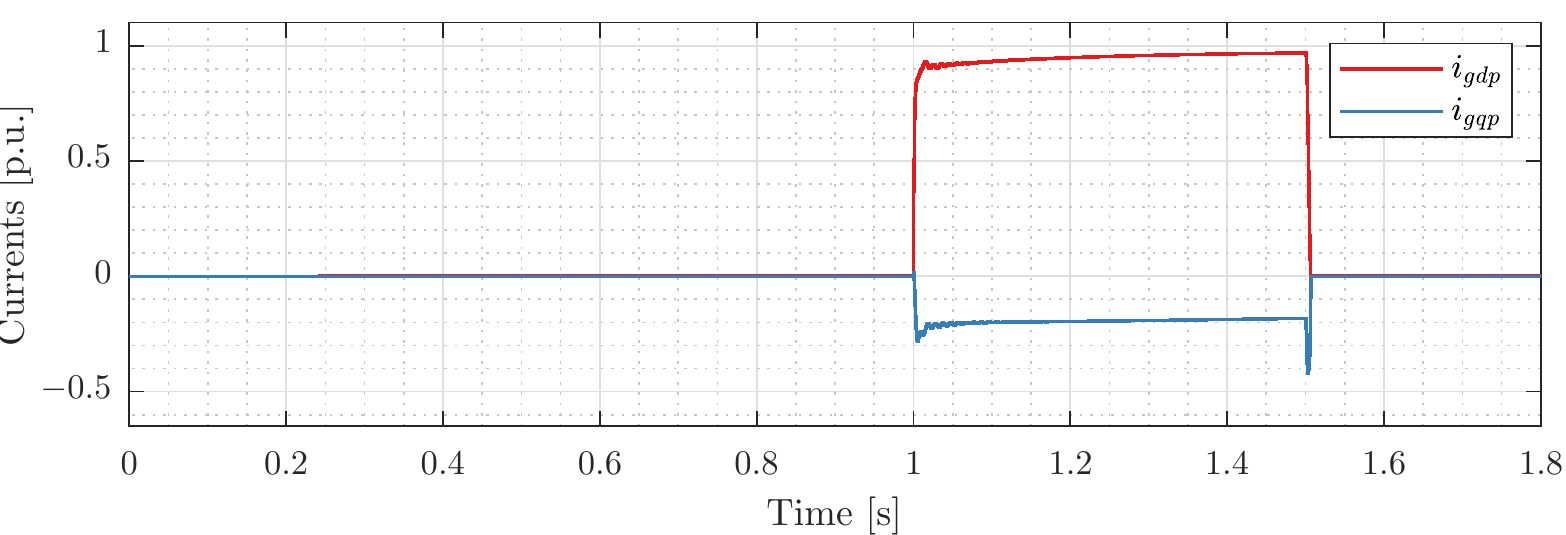}}
	\end{minipage}
	\newline
	\vspace{-0.2cm}
	\caption{{Black start scenario --- Top:  ac voltages in [p.u.] with $V_{base}^g$; Bottom: ac currents in $dq$ frame in [p.u.] with $I_{base}^g$.}} \label{fig:Blackstart}
\end{figure}

 One of the main advantages of VSC in general is the possibility to black start an ac grid when a dc energy source is available. The GFM control proposed in this paper can easily provide this functionality. In the following example, the inverter ramps up the ac voltage in $250$~ms, then, at $t=1$~s, a resistive load of $1$~p.u. is connected  and then disconnected at $t=1.5$~s. The ac voltage waveforms are shown in Fig.~\ref{fig:Blackstart} (top). When the load is connected, a small voltage drop is observed due to the effect of the virtual impedance. The current response in $dq$ frame positive sequences is presented in Fig.~\ref{fig:Blackstart} (bottom), where it is shown that the load drives current in both axes. This is due to the fact that the reference voltage $v_{qp}=0$ is located virtually in the emulated synchronous condenser.  

%
\subsection{Main grid disconnection\label{sec:GridDisc}}  
\par
\begin{figure}[t]
	\centering
	\begin{minipage}{1\columnwidth}
		\centering
		{\includegraphics[trim={0 1.1cm 0 0},clip,width=1\columnwidth]{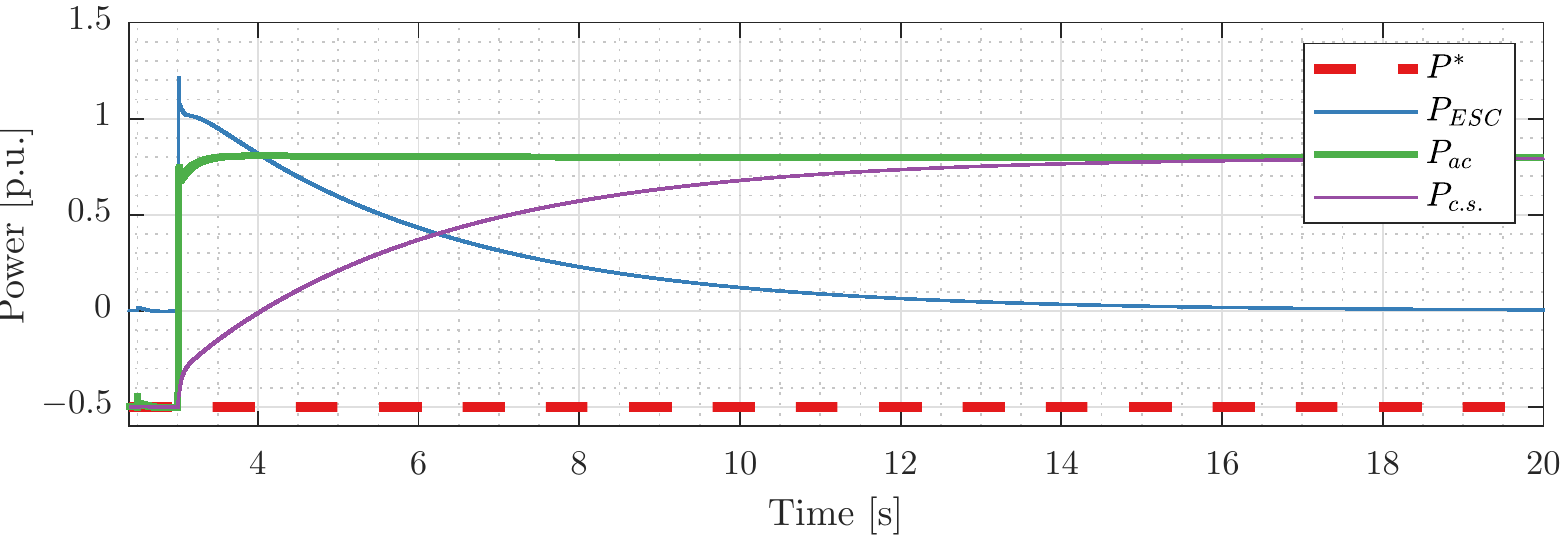}}
	\end{minipage}%
	\newline
	\begin{minipage}{1\columnwidth}
		\hfill
		{\includegraphics[width=0.99\columnwidth]{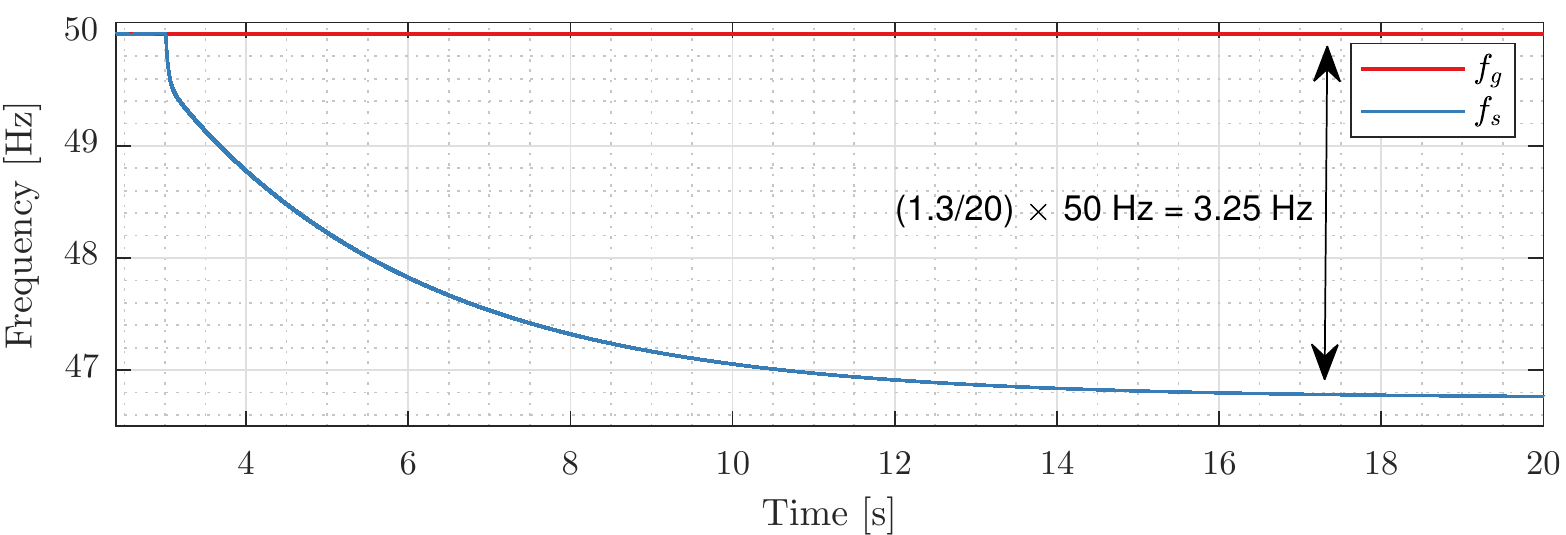}}
	\end{minipage}
	\newline
	\vspace{-0.2cm}
	\caption{{Main grid disconnection scenario --- Top:  Active powers in [p.u.] with $S_{base}$; Bottom: ac frequency in [Hz].}} \label{fig:GridDisc}
\end{figure}

 In this scenario, the battery is being charged with $P_{ac}=P_{c.s.}=-0.5$~p.u. while connected to the main grid with SCL of $400$~MVA and a load of $0.8$~p.u. connected directly at the VSC terminals. At $t=3$~s the main grid is abruptly disconnected. Active power results are shown in Fig.~\ref{fig:GridDisc} {(top)}, where it can be seen that the load connection at $t=1.5$~s disturbs the ECS. When the grid is disconnected, the power $P_{ac}$ needs to reverse from $-0.5$~p.u. to $0.8$~p.u. to fulfill the load ($\Delta P = 1.3$ p.u.). First, the ESC supplies all the load and then, progressively, the droop control action recovers. The energy provided by the ESC generates a virtual deceleration, and so the synthesized frequency $f_s$ drops down to $46.75$~Hz as dictated by the droop parameter (Fig.~\ref{fig:GridDisc}, {bottom})). {In a large interconnected or islanded grid, load-shedding mechanisms may have disconnected parts of the load to help restore the frequency before reaching such a low frequency (see \cite{GsP_Cigre}). The load-shedding thresholds depend on the grid parameters, and in small islanded grids the thresholds may be lower. This scenario shows that in case that the load-shedding mechanism fails or if the thresholds are set quite low, the BESS is capable to maintain the stability.}
%
%
%

\subsection{Fault Ride-Through Scenario\label{sec:FRT}}  
\par
\begin{figure}[t]
	\centering
	\begin{minipage}{1\columnwidth}
		\centering
		{\includegraphics[trim={0 1.1cm 0 0},clip,width=1\columnwidth]{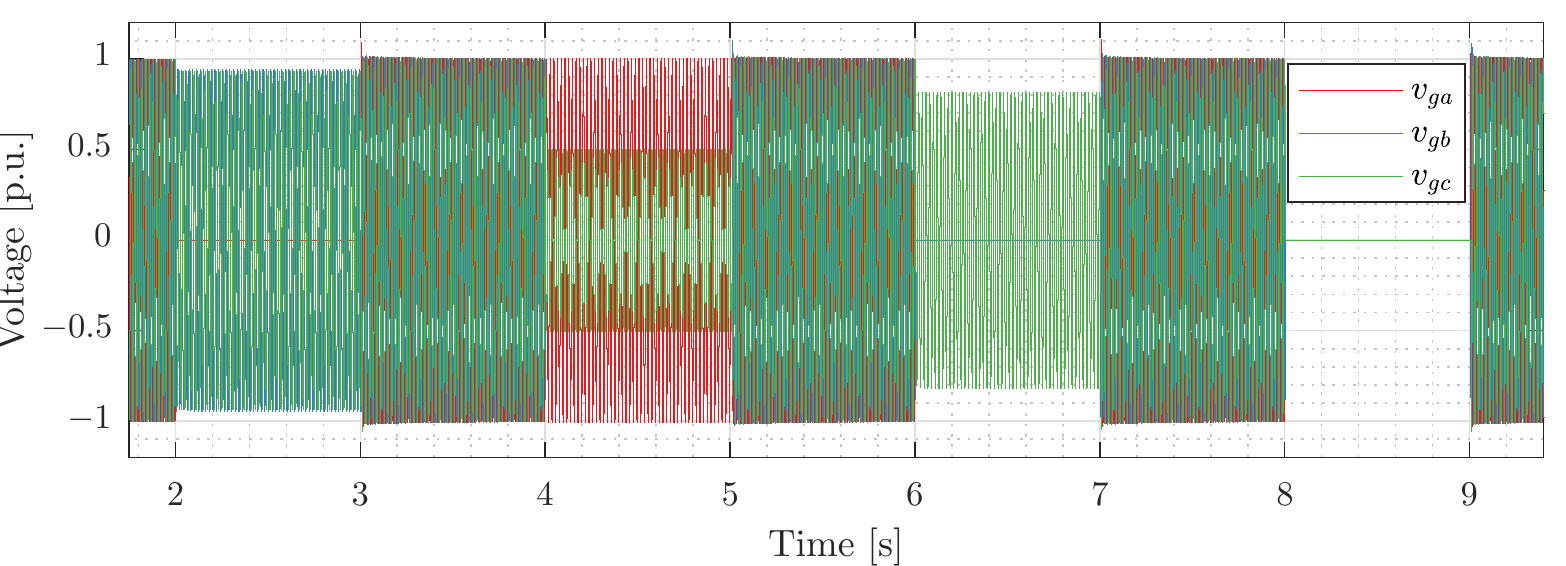}}
	\end{minipage}%
	\newline
	\begin{minipage}{1\columnwidth}
		\hfill
		{\includegraphics[width=0.99\columnwidth]{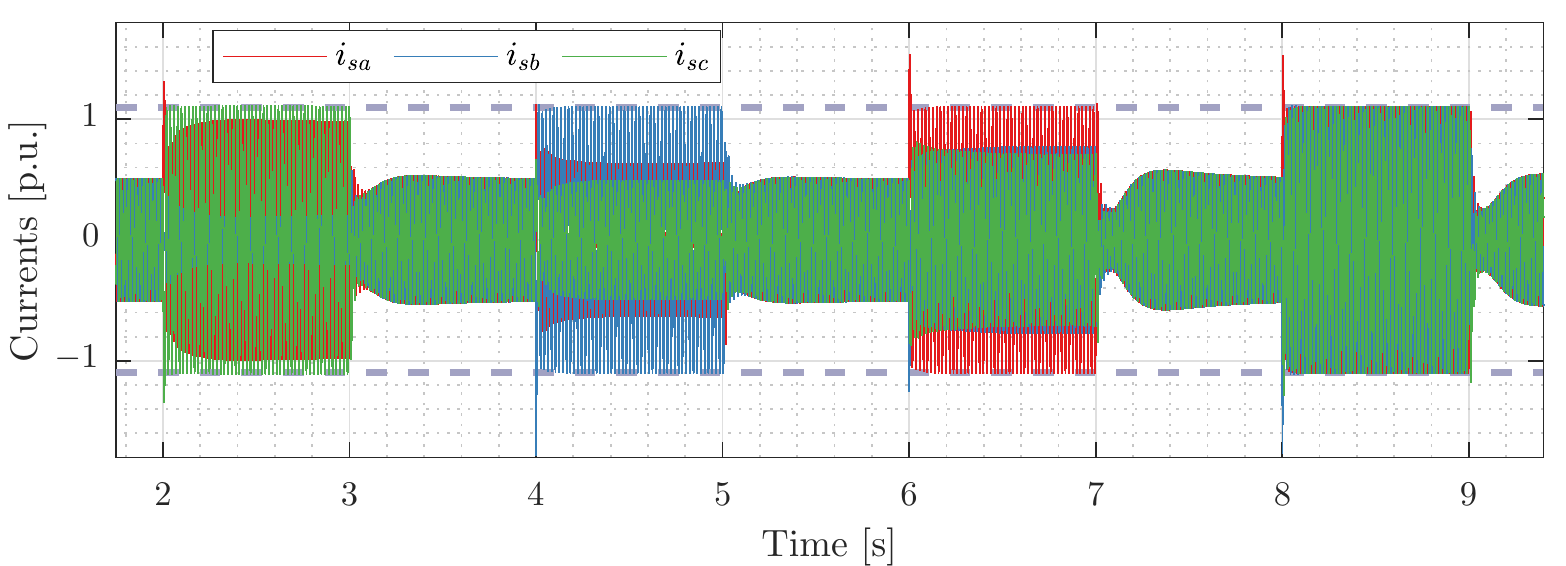}}
	\end{minipage}
	\newline
	\vspace{-0.2cm}
	\caption{{Fault ride through scenario --- Top: Grid-voltages voltages $\bm{v_{g}}$ in [p.u.] with $V_{base}^{lw}$; Bottom: Grid-side currents $\bm{i_{g}}$ in [p.u.] with {$I_{base}^{g}$}.}} \label{fig:FRTresults}
\end{figure}

 For evaluating the system during grid faults, four different faults are applied in the following scenario on the location shown in Fig.~\ref{fig:sys}: single-phase fault between time $1\:s<t<2\:s$, two-phase fault between $3\:s<t<4\:s$, two phases to ground between $5\:s<t<6\:s$ and finally three-phase fault between $7\:s<t<8\:s$. The ac grid has a SCL of $40$~MVA. Voltage and current results are shown in Fig.~\ref{fig:FRTresults}. As observed, the current is well limited to $I_{lim} =1.1$~p.u, {and the stability is maintained contrary to the Classical VSM in Section~\ref{sec:Classic_VSM}}.

%
\subsection{Extreme Phase-Jump Followed by RoCoF\label{sec:phRoc}} 
\par
\begin{figure}[t]
	\centering
	\begin{minipage}{1\columnwidth}
		\hfill
		{\includegraphics[trim={0 1.1cm 0 0},clip,width=0.985\columnwidth]{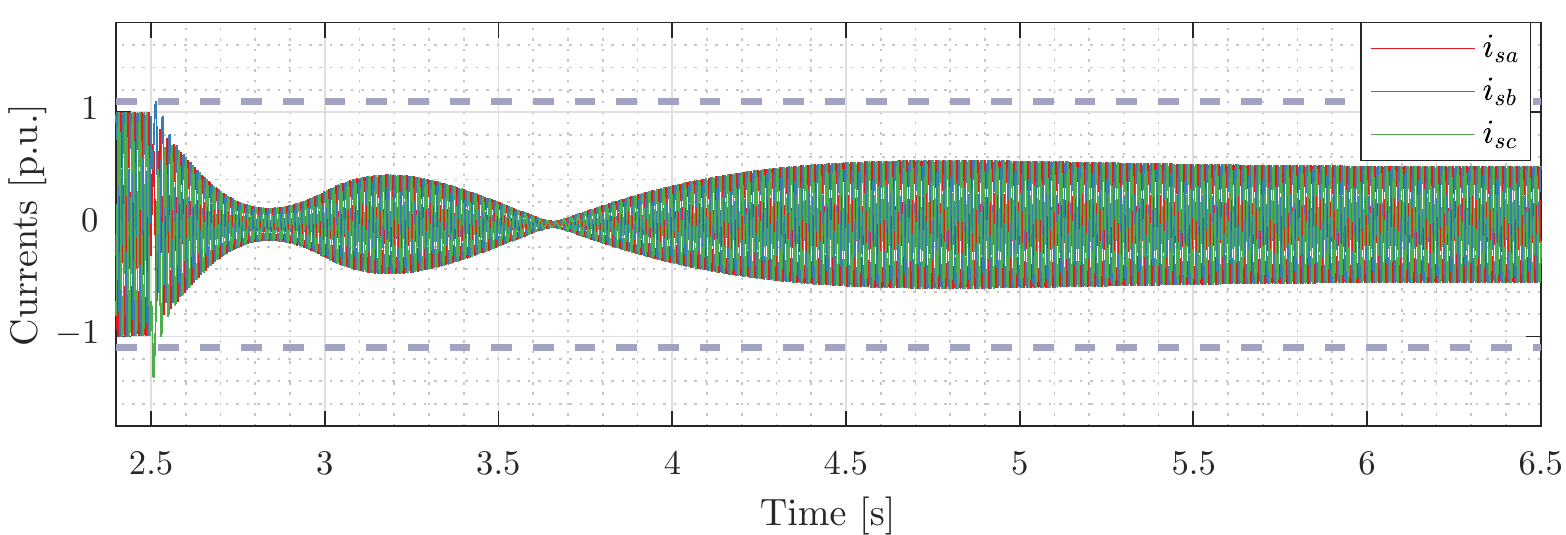}}
	\end{minipage}%
	\newline
	\begin{minipage}{1\columnwidth}
		\centering
		{\includegraphics[trim={0 1.1cm 0 0},clip,width=1\columnwidth]{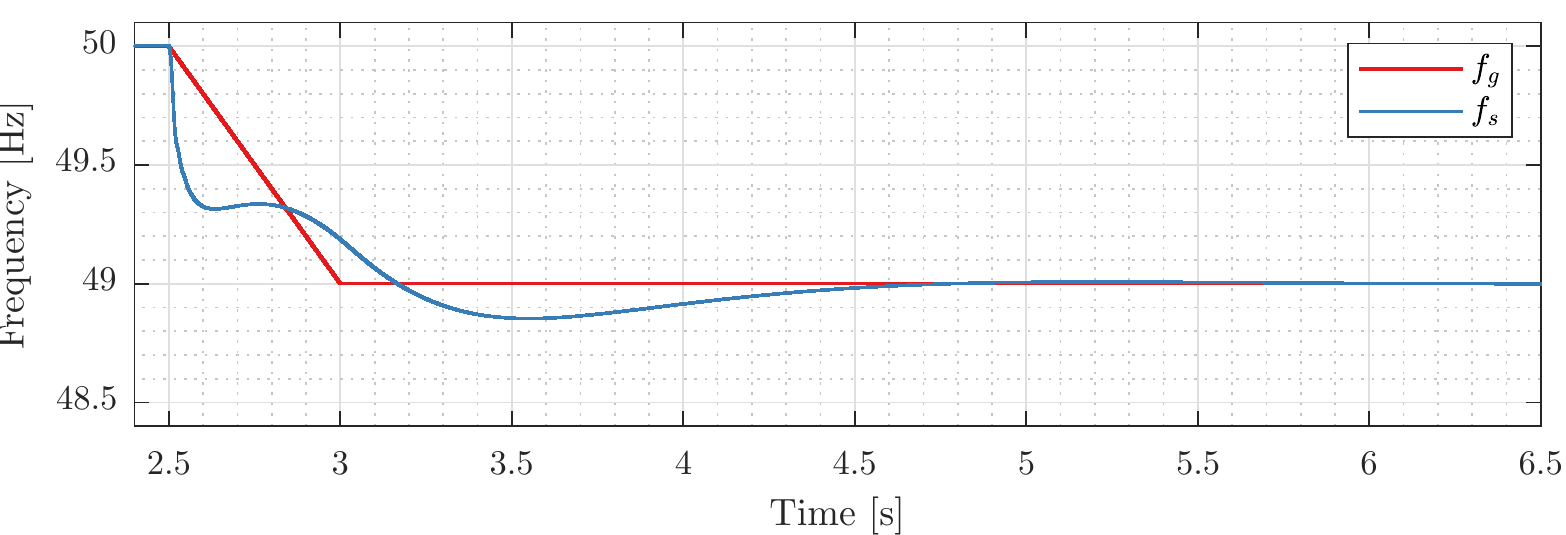}}
	\end{minipage}
		\newline
	\begin{minipage}{1\columnwidth}
		\hfill
		{\includegraphics[width=0.99\columnwidth]{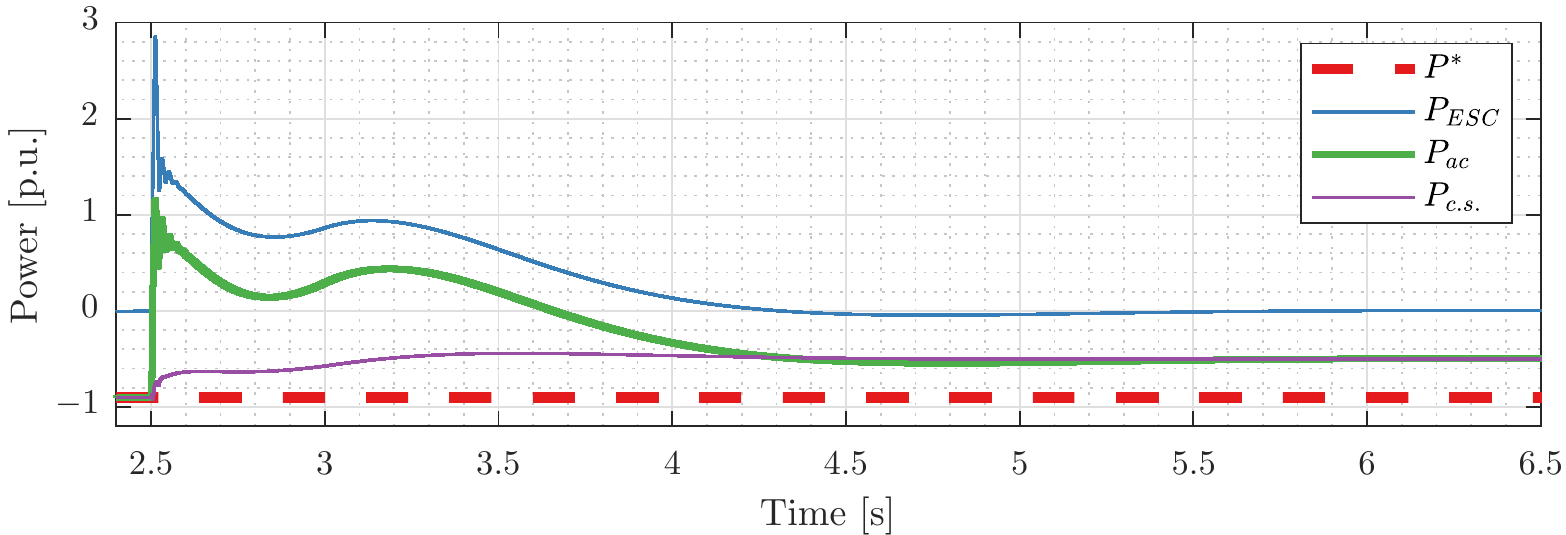}}
	\end{minipage}
	\newline
	\vspace{-0.2cm}
	\caption{{Phase-Jump of $-80$~deg. followed by RoCoF while charging --- Top:  ac currents in [p.u.] with $I_{base}^s$; Middle: ac frequency in [Hz]; Bottom:  active powers in [p.u.] with $S_{base}$.}} \label{fig:phjump_charge}
\end{figure}

\begin{figure}[t]
	\centering
	\begin{minipage}{1\columnwidth}
		\hfill
		{\includegraphics[trim={0 1.1cm 0 0},clip,width=0.985\columnwidth]{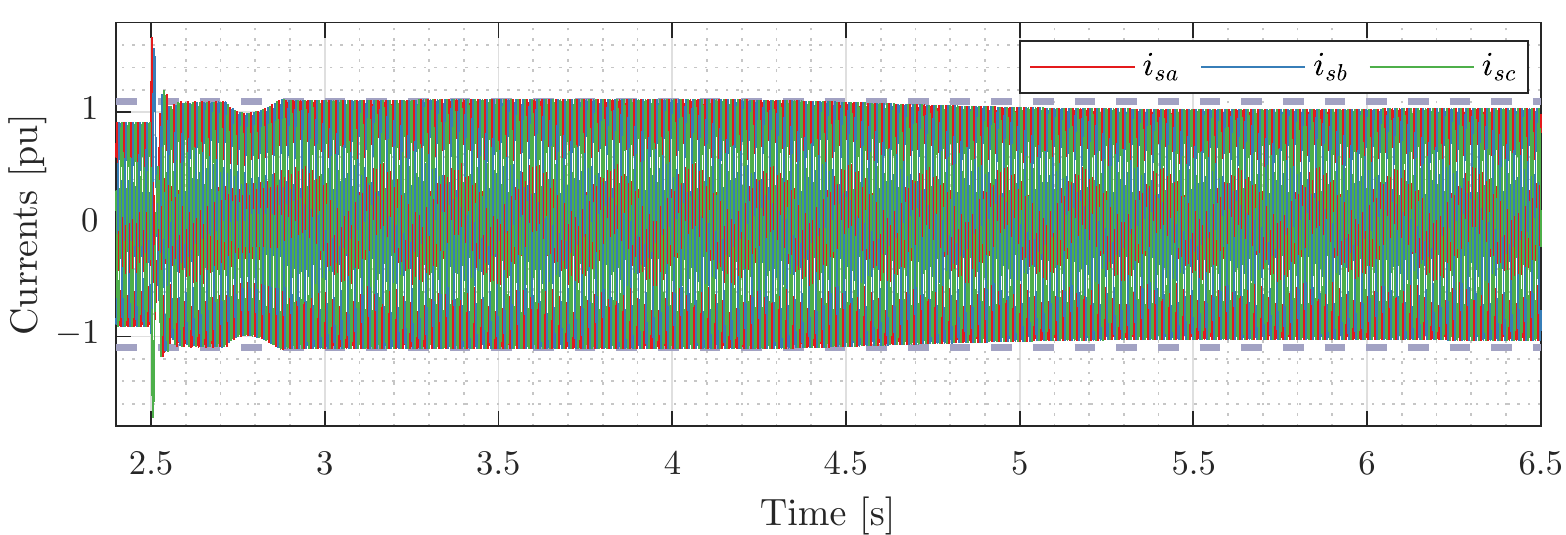}}
	\end{minipage}%
	\newline
	\begin{minipage}{1\columnwidth}
		\centering
		{\includegraphics[trim={0 1.1cm 0 0},clip,width=1\columnwidth]{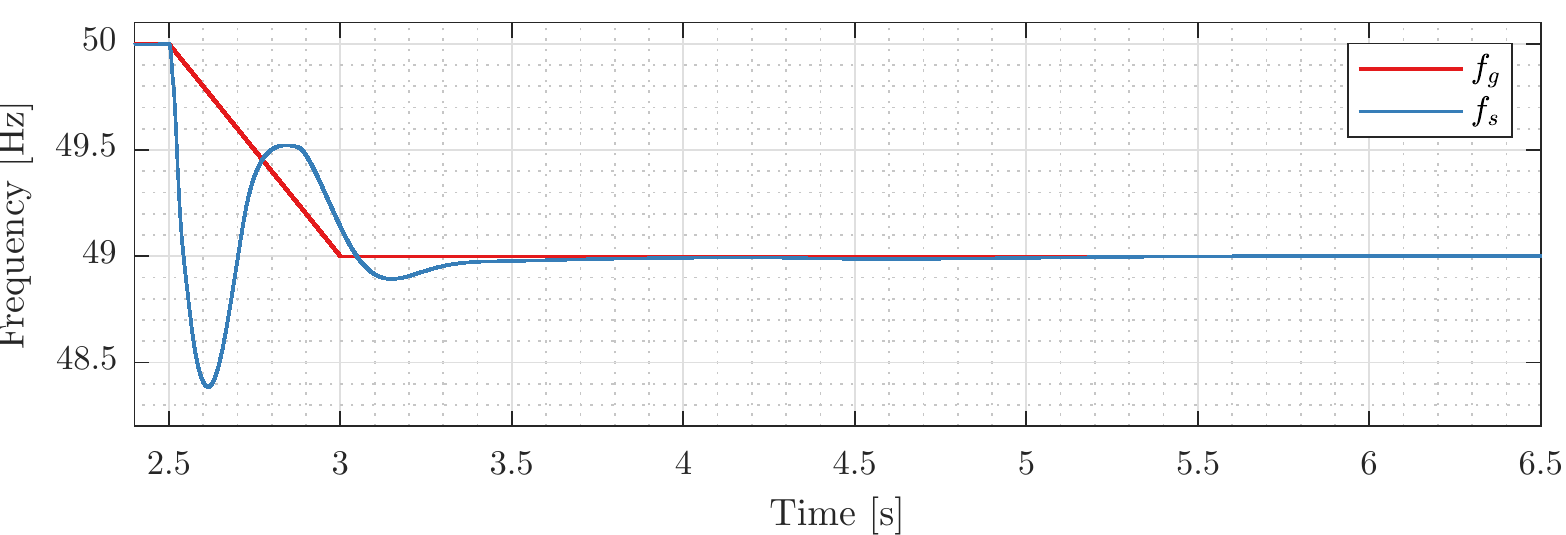}}
	\end{minipage}
	\newline
	\begin{minipage}{1\columnwidth}
		\hfill
		{\includegraphics[width=0.99\columnwidth]{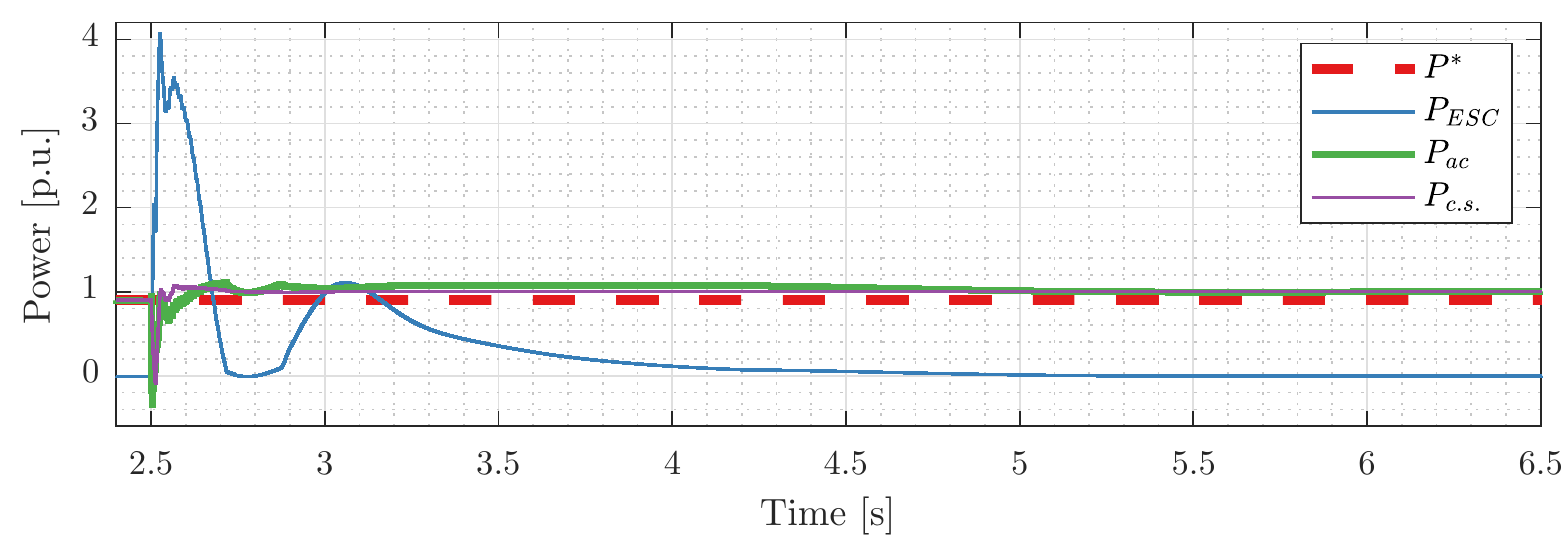}}
	\end{minipage}
	\newline
	\vspace{-0.2cm}
	\caption{{Phase-Jump of $-80$~deg. followed by RoCoF while discharging --- Top:  ac currents in [p.u.] with $I_{base}^s$; Middle: ac frequency in [Hz]; Bottom:  active powers in [p.u.] with $S_{base}$.}} \label{fig:phjump_discharge}
\end{figure}

 The last scenario considers an ac grid with SCL = $4$~MVA, and consists in an extreme phase jump of $-80\degree$ in addition to a frequency slew of $2$~Hz/s \cite{KIT_VSM_FRT, ESIGgeneral, MITSUBISHI_FACTS_GFM}. Results of the simulated scenario when the BESS is charging ($P_{ac}=-0.9$~p.u.) are shown in Fig.~\ref{fig:phjump_charge}; while Fig.~\ref{fig:phjump_discharge} show the results when the BESS is discharging ($P_{ac}=0.9$~p.u.). 

\par In both cases, the phase jump of $-80\degree$ entails a severe instantaneous active power demand to be injected in the grid: when the BESS is charging, the headroom of active power is $2$~p.u., meaning that an instantaneous full power inversion is demanded. As shown in Fig.~\ref{fig:phjump_charge}, the control strategy is capable of overcoming this scenario, but it is of utmost importance to verify that the primary power source can withstand this fast power inversion. When the power of the VSC is already at $P_{ac}=1$~p.u. as in Fig.~\ref{fig:phjump_discharge}, the current is limited to the maximum value and there is not much room for overpower. Since the current is saturated, the virtual power of the ESC, $P_{ESC}$, ramps quickly to $4$~p.u.: since this power is virtual, it is not physically flowing through the inverter, so there is no risk of damaging the equipment. Frequency results for both cases are shown in Fig.~\ref{fig:phjump_charge} for charging and Fig.~\ref{fig:phjump_discharge} for discharging (middle). For the first case, the frequency drop of $f_s$ is not as important as for the latter case due to the current limitation action. Nevertheless, both scenarios are surpassed satisfactorily. 

%
\vspace{-0.25cm}
\section{Conclusion} \label{sec:Conclusions}
%
%
\par For the massive insertion of inverter-based generation in power systems, there is a general consensus that Grid-Forming converters are likely to play a very important role. From a control perspective, it is necessary to develop strategies that are capable of withstanding severe grid conditions. This paper presents a GFM control algorithm which combines and complements a virtual current source and an Emulated Synchronous Condenser. The current source allows a fast power reference tracking response and system services such as primary frequency control, while the ESC in parallel provides the GFM characteristic and inertial effect. The strategy is evaluated via time-domain simulations of a 2-MW BESS, showing satisfactory results for varied conditions without the need for extra control logics. {Ongoing research and future works aim at evaluating this strategy with more detailed grids, including a mix of generators and other converters to analyse the interactions in microgrids and larger power systems.}

%
\bibliography{Biblio_GridForming}
\bibliographystyle{ieeetr} 
%
\end{document}